\documentclass[11pt]{article}

\usepackage[english]{babel}
\usepackage{sao1}
\usepackage{graphics}
\usepackage{epsfig}

\begin{document}

\title{X-ray emission and the incidence of magnetic fields in the massive stars of the Orion Nebula Cluster}
\author{V. Petit\inst{1,2} \and G.A. Wade\inst{2} \and T. Montmerle\inst{3}
	\and Drissen\inst{1} \and N. Grosso\inst{3} \and F. Menard\inst{3}}
\institute{D\'epartement de Physique, de G\'enie Physique et d'Optique and Observatoire du mont M\'egantic, Universit\'e Laval, Qu\'ebec, QC G1K 7P4, Canada	    \and Department of Physics, Royal Military College of Canada, PO Box 17000, Station `Forces', Kingston, Ontario, K7K 4B4, Canada 
	    \and Laboratoire d'Astrophysique de Grenoble, Universit\'e Joseph-Fourier, F-38041 Grenoble, France}

\maketitle 

\begin{abstract}

Magnetic fields have been frequently invoked as a likely source of variability and confinement of the winds of 
massive stars. To date, the only magnetic field detected in O-type stars are those of $\theta^1$ Ori C (HD 37022; 
Donati et al. 2002), the brightest and most massive member of the Orion Nebula Cluster (ONC), and HD 
191612 (Donati et al. 2006). Notably, $\theta^1$ Ori C is an intense X-ray emitter, and the source of these X-rays 
is thought to be strong shocks occurring in its magnetically-confined wind (Babel \& Montmerle 1997a, Donati 
et al. 2002). 

Recently, Stelzer et al. (2005) found significant X-ray emission from all massive stars in the ONC. Periodic 
rotational modulation in X-rays and other indicators suggested that $\theta^1$ Ori C may be but one of many magnetic 
B- and O-type stars in this star-forming region. In 2005B we carried out sensitive ESPaDOnS observations 
to search for direct evidence of such fields, detecting unambiguous Zeeman signatures in two objects. 

\keywords{stars: early-type $-$ stars: magnetic fields $-$ stars: mass-loss}
\end{abstract}

\section{Introduction}

The existence of magnetic fields in early-type stars remains a mystery. The intermediate-mass magnetic Ap-Bp stars, with observed $\sim$kG dipolar fields, are well known, but because they have radiative envelopes, they cannot host the dynamo-generated surface magnetic fields found in late-type convective stars like the Sun. These fields are believed to be fossil remnants of either interstellar magnetic fields swept up during the star formation process or fields produced by a pre-main sequence envelope dynamo that has since turned off. 

In more massive stars, magnetic fields have only been discovered recently, mostly via clues provided by unusual X-ray properties. Traditionally, the X-ray emission from O and B stars, with a typical level $L_X/L_{bol} \sim 10^{-7}$, has been explained by radiative instabilities, via a multitude of shocks in the wind (Lucy \& White 1980, Owocki \& Cohen 1999). However, the very strong and rotationally modulated X-ray emission of the brightest Trapezium star, $\theta^{1}$ Ori C (O7, P=15.4~d), was explained by Babel \& Montmerle (1997a, 1997b) in terms of the ``magnetically confined wind shock'' model (MWCS). In this model, the stellar magnetic field is sufficiently strong, and the radiative wind sufficiently weak, to allow a dipolar magnetic field to confine the outflowing wind in the immediate circumstellar environment, resulting in a closed magnetosphere with a large-scale equatorial shock which heats the wind plasma. In this way, the X-ray emission is enhanced and may be modulated by stellar rotation. Babel and Montmerle thereby predicted the existence of a dipolar magnetic field in $\theta^{1}$ Ori C; such a field was subsequently discovered by Donati et al. (2002). One can therefore conclude that an unusually high, and/or rotationally modulated X-ray emission, is a strong indication of the presence of closed magnetospheres in massive stars.

An exciting application of these results has been provided by the ``Chandra Orion Ultradeep Project'' (COUP). This program consisted of a 9.7 day almost continuous observation with the ACIS detector (FOV 17' $\times$ 17'), centered on the Orion Nebula Cluster (ONC). In all, 1616 sources were detected, characterized and identified (Getman et al. 2005). The COUP study of the OBA population (20 stars) by Stelzer et al. (2005) was aimed at disentangling the respective roles of wind and magnetic fields in X-ray emission of these stars. X-ray emission by standard radiative shocks is found to be the dominant mechanism for the subsample of 9 high-mass O to early-B stars which have strong winds. However, 3 of the high-mass stars showed X-ray properties suggestive of the presence of magnetic fields: $\theta^{1}$ Ori C (O7; discussed above), $\theta^{1}$ Ori A (B0), and JW660 (B3). On the other hand, a fourth star, Par 1772 (B2), a tentative helium-strong star, showed only a very weak X-ray emission.
We have undertaken a study with ESPaDOnS to explore the role of magnetic fields in producing this diversity of X-ray behaviours.

\section{Observations}

In January 2006 we conducted circular polarisation observations with ESPaDOnS at CFHT to search for direct evidence of magnetic fields in the ONC massive stars. We obtained single high S/N Stokes $V$ spectra of 8 of the massive OB stars. The mean Stokes I and V profiles were extracted with the Least Square Deconvolution technique (LSD) of Donati et al. (1997), which allows the use of many lines to increase the level of detection of a magnetic field Stokes V signature. The LSD line masks were carefully chosen, in order to maximize the resultant LSD S/N. Several experiments were attempted, including the removal of contaminated lines and helium lines, along with comparison with synthetic spectra.

\begin{figure}
\centerline{\epsfig{file=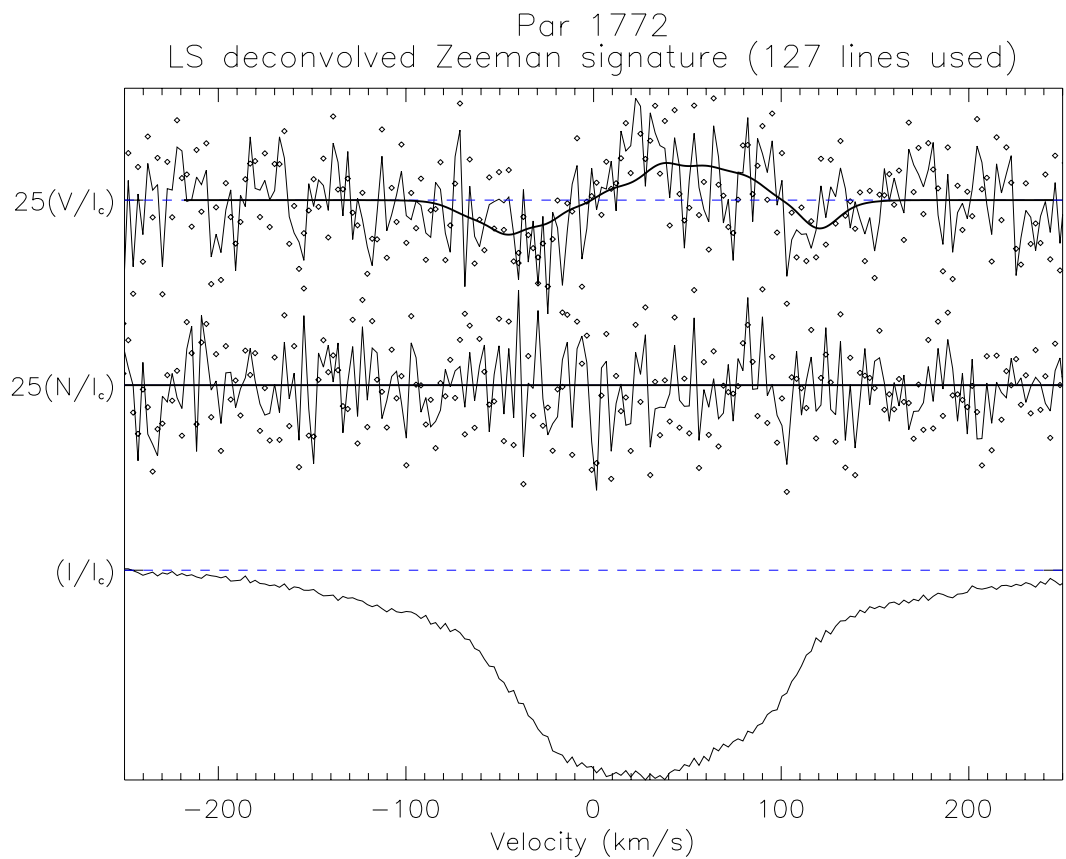,width=14cm}}
\caption{\label{lsd} {\bf Least Square 
Deconvolved Stokes I and V profiles.} The thin lines show the mean Stokes I profile (bottom), 
the mean Stokes V profile (top) and the N diagnostic null profile (middle). The thick line is the 
best dipole model fit from our magnetic analysis.}
\end{figure}

We detected clear magnetic (Stokes $V$) signatures in spectra of 2 stars: $\theta^{1}$ Ori C, which was already known to be magnetic, and Par 1772, a new detection (Fig. \ref{lsd}). For the remaining 6 stars, we observed no Stokes $V$ detection. Table \ref{tobs} gives a summary of the target properties, along with the longitudinal field error bar that we achieved. 

\begin{table}[h]
\caption{\label{tobs}Target information and longitudinal field error bars achieved.}
\vspace{2ex}
\begin{center}
\begin{tabular}{lccccc}\hline
    & V  & Type & vsin$i$ (km/s) & Obs. Time & $\sigma_B$ (G)\\
\hline
{\bf$\theta^{1}$ Ori C} & 5.1 & O7      & 50    & 0h53 & 37\\
$\theta^{2}$ Ori A        & 5.1 & O9.5   & 110  & 1h20 & 346\\
$\theta^{1}$ Ori A        & 6.7 & B0   	& 115  & 2h40 & 225\\
$\theta^{1}$ Ori D        & 6.7 & B0.5	& 50    & 2h40 & 30\\
NU Ori 		            & 6.8 & B1 	& 160  & 2h40 & 116\\
$\theta^{2}$ Ori B        & 6.0 & B1 	& 25    & 0h40 & 12\\
{\bf Par 1772}    	  & 8.4 & B2       & 120  & 2h40 & 72\\
JW 660		            & 9.7 & B3       & 210  & 4h25 & 460\\
\hline
\end{tabular}
\end{center}
\end{table}

\section{Magnetic analysis}

To constrain admissible surface magnetic field configurations of these stars, we have compared the observed Stokes $V$ spectra with synthetic profiles computed assuming a large grid of dipolar magnetic field models. Those models are characterized by different values of the magnetic tilt angle $\beta$ and dipolar intensity $B_{\rm d}$. 

\begin{figure}
\centerline{\epsfig{file=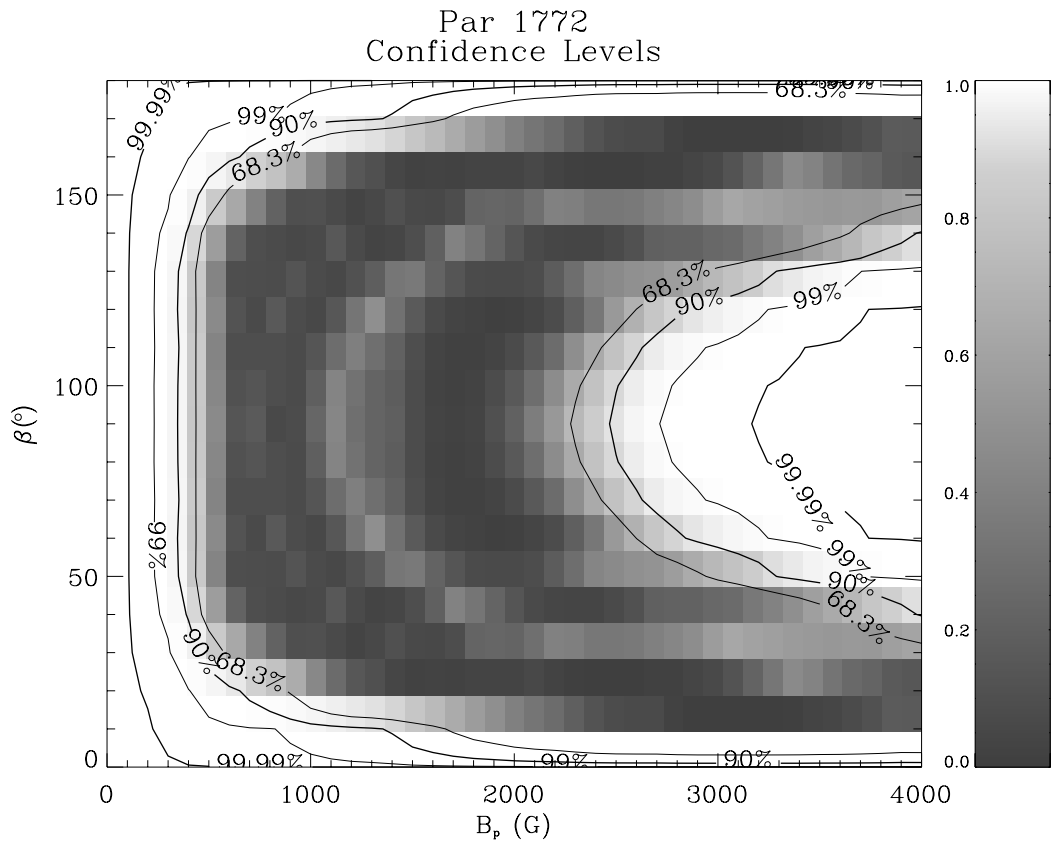,width=14cm}}
\caption{\label{zeeman} {\bf Confidence map for admissible dipole 
field strength B$_p$ and obliquity $\beta$} . There is 99\% probability that the real surface field strength 
and obliquity are within the 99\% contour. From this map, constructed from a single observation, 
we can say that the field is stronger than 400~G (at 99\% confidence), but more observations are 
required to significantly constrain the dipole inclination and the field strength upper limit. 
}
\end{figure}

By computing the reduced $\chi^2$ of the model with respect to the observations, we have constructed confidence maps of admissible (and rejectable) field configurations for each star. One such map is illustrated in Fig. \ref{zeeman} for Par 1772. Our single observation of this star constrains the surface magnetic field to be above 400~G with 99\% confidence. However, we have weak constraint on the dipole tilt and on the maximum field strength. This behaviour results from the limited information afforded by a single observation.

Analogous maps have been obtained for all observed stars. For those in which no Stokes $V$ detection is obtained, our maps allow us in principle to determine upper limits on admissible field configurations. 
This is crucial: a magnetic field of specific minimum strength is required to confine a wind of known mass flux. Hence if we can determine an upper limit on admissible magnetic fields of the undetected stars, we can potentially rule out magnetically-confined wind shocks as the source of their X-ray emission and variability. 
In addition, these upper limits provide unique data with which to confront models of magnetic field origin in neutron stars and magnetars, such as that proposed by Ferrario \& Wickramasinghe (2006). 
Unfortunately, our confidence maps of the undetected stars suffer from the same characteristics as the map of Par 1772: the maximum admissible field strength (upper limit) is poorly constrained. Because when $\beta$ = 90¡$-i$, geometries exist for which there is no Stokes V signature, notwithstanding the field strength. 
Fortunately, our numerical experiments show that a by obtaining at least one additional Stokes $V$ observation of each star, we can constrain significantly the orientation of the dipole, leading to a dramatic reduction in the range of probable field strengths by a full order of magnitude.  Nevertheless, we can still give a rough estimate of the magnetic field, by taking reasonable values for $\beta$ and $i$ (ie, by ignoring the degenerate configuration). Those first estimates are given in Table \ref{tb}. A more robust analysis is currently underway, along with the planning of multi-phase spectropolarimetric observations.

\begin{table}[h]
\caption{\label{tb}{\bf First estimates of the field strengths and upper limits for the ONC massive stars.} Calculated from 1$\sigma$ contours at $\beta=90\degr$.}
\vspace{2ex}
\begin{center}
\begin{tabular}{ll}\hline
	&	Bd (G) \\ \hline
$\theta^1$ Ori C	&	1100$\pm$100 (3$\sigma$)\\
$\theta^2$ Ori A	&	$<$600 \\
$\theta^1$ Ori A	&	$<$300 \\
$\theta^1$ Ori D	&	$<$150 \\
NU Ori			&	TBD\\
$\theta^2$ Ori B	&	$<$150 \\
Par 1772			&	800$-$2500 \\
JW 660			&	$<$4000 \\
\hline
\end{tabular}
\end{center}
\end{table}

\section{Conclusion}

Using Stokes V mesurements obtained with ESPaDOnS spectropolarimeter at CFHT, we were able to detect the presence of a new magnetic star in the Orion Nebula Cluster. Par 1772 has a dipole field component greater than 400~G (at 99\% confidence). However, the precise magnetic configuration cannot be fully constrained with a single observation; neither can the maximum field strengh in the non-detection cases. However, a more detail analysis, combined with multi-phase observations will enable more precise estimates.
Once completed, this study of the Orion stellar cluster will represents a {\bf complete} magnetic survey of a {\bf co-evolved} and {\bf co-environmental} population of massive stars. 

A important issue of this study regards magnetic fields in even more massive stars. It is related to neutron star population studies. Ferrario \& Wickramasinghe (2006) explored the hypothesis that neutron star magnetic fields are of fossil origin, and predicted a field distribution for the OB star progenitors (above 8 solar masses). According to their model, $\sim$7\% of the main sequence OB stars should have a magnetic field in excess of 1kG. Interestingly, we find 2 stars (25\%) with fields above 1 kG in the ONC. As the fossil hypothesis depends on the environment (in the case of a interstellar magnetic fields swept up), it could be possible that the Orion Nebular Cluster is overly magnetized. Similar studies of other young OB clusters would therefore be really interesting. On the other hand, if the magnetic field origin is intrinsic to the star itself (envelope dynamo remnant coming from a pre-main sequence phase for exemple) the fraction of magnetized stars should be similar in all OB clusters.

\begin{acknowledgements}

\end{acknowledgements}


\begin{thebibliography}{}
\bibitem{}
	Lucy, L. B. \& White, R. L. 1980, ApJ {\bf241}, 300
\bibitem{}
	Owocki, S. P. \& Cohen D. H. 1999, ApJ {\bf520}, 833
\bibitem{}
	Babel, J., \& Montmerle, T. 1997a, A\&A {\bf323}, 121
\bibitem{}
	Babel, J., \& Montmerle, T. 1997b, ApJ {\bf485}, L29
\bibitem{}
	Donati, J.-F., Babel, J., Harries, T. J., Howarth, I. D., Petit, P. \& Semel, M. 2002, MNRAS {\bf333}, 55
\bibitem{}
	Getman, K. V., Feigelson, E. D., Grosso, N., McCaughrean, M. J., Micela, G., Broos, P., Garmire, G. \& Townsley, L. 2005, ApJS {\bf160}, 353
\bibitem{}
	Stelzer, B., Flaccomio, E., Montmerle, T., Micela, G., Sciortino, S., Favata, F., Preibisch, T. \& Feigelson, E. D. ApJS {\bf160}, 557
\bibitem{}
	Ferrario, L. \& Wickramasinghe, D. T. 2006, MNRAS {\bf367}, 1323
\bibitem{}
	Donati, J.-F., Semel, M., Carter, B. D., Rees, D. E. \& Collier Cameron, A. 1997, MNRAS {\bf291}, 658
\bibitem{}
	Donati, J.-F., Howarth, I. D., Bouret, J.-C., Petit, P., Catala, C. \& Landstreet, J. 2006, MNRAS {\bf365}, L6

\end{thebibliography}
\end{document}